\documentclass[12pt]{article}
\usepackage{pdproc,epsfig} 

\begin{document}

\newcommand{\beq}{\begin{equation}}
\newcommand{\eeq}{\end{equation}}
\newcommand{\bea}{\begin{eqnarray}}
\newcommand{\eea}{\end{eqnarray}}
\def\simlt{\buildrel < \over {_{\sim}}}
\def\simgt{\buildrel > \over {_{\sim}}}
\newcommand{\ttbs}{\char'134}
\newcommand{\AmS}{{\protect\the\textfont2
  A\kern-.1667em\lower.5ex\hbox{M}\kern-.125emS}}
\renewcommand{\thefootnote}{\alph{footnote}}

\title{NEUTRINO MASS, FLAVOUR AND CP VIOLATION}

\author{S.F.KING}

\address{Department of Physics and Astronomy, \\ 
        University of Southampton, Southampton SO17 1BJ, U.K.
 {\rm E-mail: sfk@hep.phys.soton.ac.uk}}

\abstract{In this talk I discuss the connection between neutrino mass,
flavour and CP violation.
I focus on three neutrino
patterns of neutrino masses and mixing angles,
and the corresponding Majorana mass matrices.
I discuss the see-saw mechanism, and show how it may be applied
in a very natural way to give a neutrino mass hierarchy with
large atmospheric and solar angles by assuming 
sequential right-handed neutrino dominance.
I then distinguish between heavy sequential dominance and
light sequential dominance, and show how lepton flavour
violation in the CMSSM provides a way to discriminate between
these two possibilities. I also show that for a well motivated class
of light sequential dominance models there is a link between
leptogenesis and CP violation measurable in neutrino oscillation experiments. }
   
\normalsize\baselineskip=15pt

\section{NEUTRINO MASSES AND MIXING ANGLES}

There is by now strong evidence for neutrino oscillations in both the
atmospheric and solar neutrino sectors 
\cite{bahcall,smirnov,mcdonald,asuzuki,kajita,ysuzuki,fogli}.
The minimal neutrino sector required to account for the
atmospheric and solar neutrino oscillation data consists of
three light physical neutrinos with left-handed flavour eigenstates,
$\nu_e$, $\nu_\mu$, and $\nu_\tau$, defined to be those states
that share the same electroweak doublet as the left-handed
charged lepton mass eigenstates.
Within the framework of three--neutrino oscillations,
the neutrino flavor eigenstates $\nu_e$, $\nu_\mu$, and $\nu_\tau$ are
related to the neutrino mass eigenstates $\nu_1$, $\nu_2$, and $\nu_3$
with mass $m_1$, $m_2$, and $m_3$, respectively, by a $3\times3$ 
unitary matrix $U$
\cite{pontecorvo,Maki:1962mu,Lee:1977qz}
\begin{equation}
\left(\begin{array}{c} \nu_e \\ \nu_\mu \\ \nu_\tau \end{array} \\ \right)=
\left(\begin{array}{ccc}
U_{e1} & U_{e2} & U_{e3} \\
U_{\mu1} & U_{\mu2} & U_{\mu3} \\
U_{\tau1} & U_{\tau2} & U_{\tau3} \\
\end{array}\right)
\left(\begin{array}{c} \nu_1 \\ \nu_2 \\ \nu_3 \end{array} \\ \right)
\; .
\end{equation}

Assuming the light neutrinos are Majorana,
$U$ can be parameterized in terms of three mixing angles
$\theta_{ij}$ and three complex phases $\delta_{ij}$.
A unitary matrix has six phases but three of them are removed 
by the phase symmetry of the charged lepton Dirac masses.
Since the neutrino masses are Majorana there is no additional
phase symmetry associated with them, unlike the case of quark
mixing where a further two phases may be removed.
The neutrino mixing matrix may be parametrised by 
a product of three complex Euler rotations,
\begin{equation}
U=U_{23}U_{13}U_{12}
\end{equation}
where
\begin{equation}
U_{23}=
\left(\begin{array}{ccc}
1 & 0 & 0 \\
0 & c_{23} & s_{23}e^{-i\delta_{23}} \\
0 & -s_{23}e^{i\delta_{23}} & c_{23} \\
\end{array}\right)
\end{equation}

\begin{equation}
U_{13}=
\left(\begin{array}{ccc}
c_{13} & 0 & s_{13}e^{-i\delta_{13}} \\
0 & 1 & 0 \\
-s_{13}e^{i\delta_{13}} & 0 & c_{13} \\
\end{array}\right)
\end{equation}

\begin{equation}
U_{12}=
\left(\begin{array}{ccc}
c_{12} & s_{12}e^{-i\delta_{12}} & 0 \\
-s_{12}e^{i\delta_{12}} & c_{12} & 0\\
0 & 0 & 1 \\
\end{array}\right)
\end{equation}

where $c_{ij} = \cos\theta_{ij}$ and $s_{ij} = \sin\theta_{ij}$.
Note that the allowed range of the angles is
$0\leq \theta_{ij} \leq \pi/2$. 
Since we have assumed that the neutrinos are Majorana, 
there are two extra phases, but only one combination
$\delta = \delta_{13}-\delta_{23}-\delta_{12}$
affects oscillations.

\begin{figure}
\vspace*{13pt}
        \mbox{\epsfig{figure=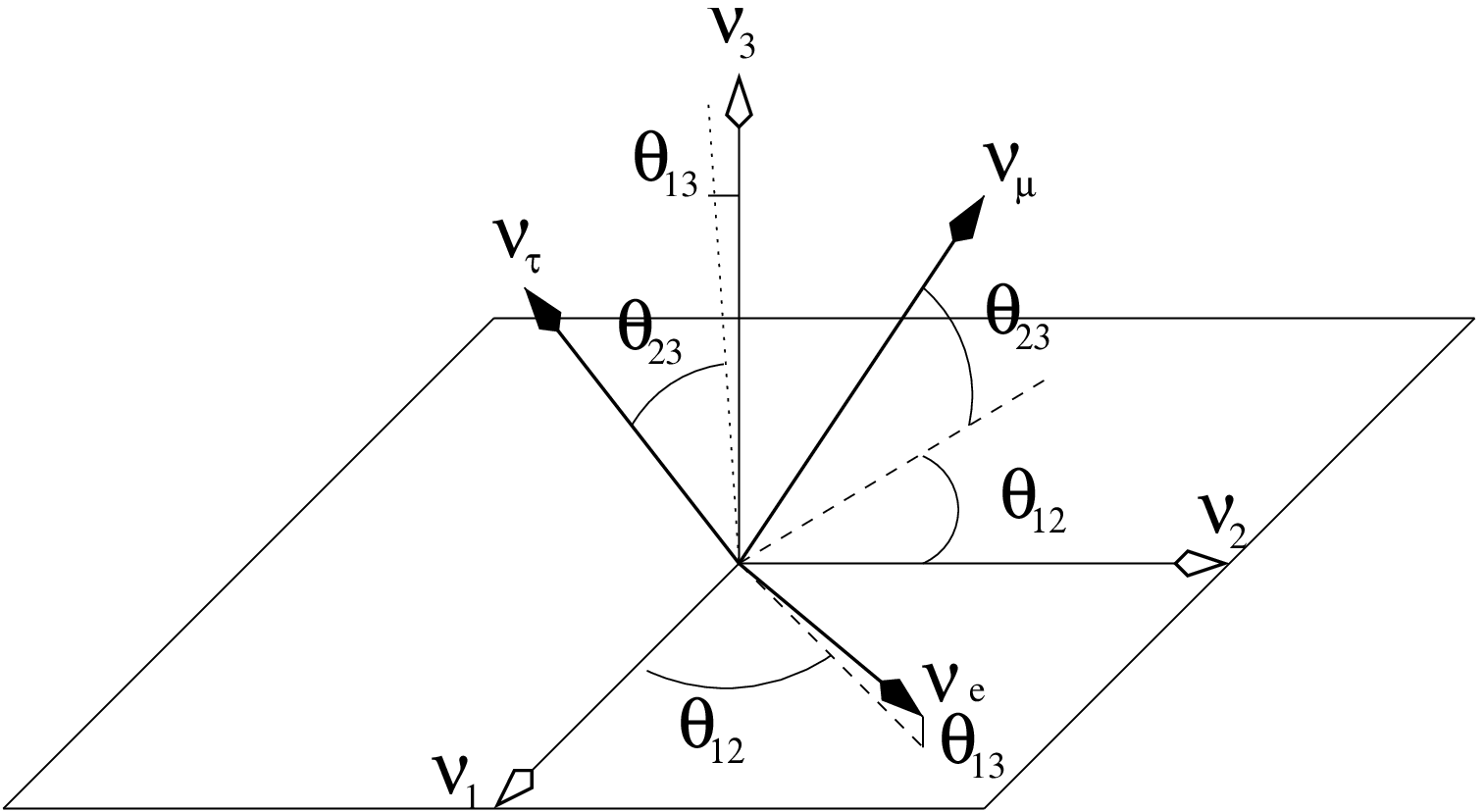,width=16.0cm}}
\vspace*{1.4truein}		
\caption{A graphical illustration of the neutrino mixing angles.
This figure ignores the phases
so that the neutrino mixing matrix is constructed
as a product of three Euler rotations $U=R_{23}R_{13}R_{12}$.
The atmospheric angle is $\theta_{23} \approx \pi/4$, 
the CHOOZ angle is $\theta_{13} \simlt 0.2$, and the solar angle is
$\theta_{12} \approx \pi/6$.}
\label{angles}
\end{figure}

Ignoring phases, the relation between 
the neutrino flavor eigenstates $\nu_e$, $\nu_\mu$, and $\nu_\tau$ and
the neutrino mass eigenstates $\nu_1$, $\nu_2$, and $\nu_3$
is just given as a product of three Euler rotations as
depicted in Fig.\ref{angles}.

\begin{figure}
\vspace*{13pt}
        \mbox{\epsfig{figure=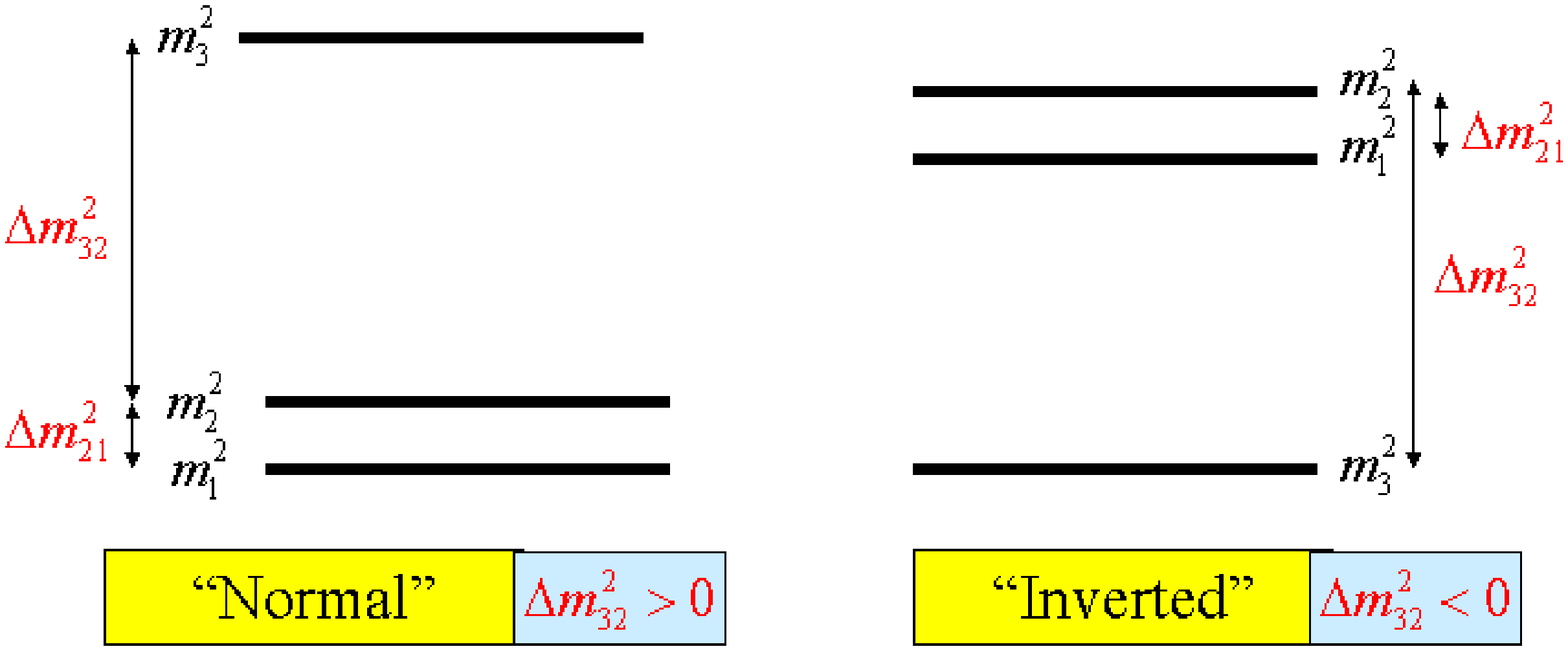,width=16.0cm}}
\vspace*{1.4truein}		
\caption{Alternative neutrino mass 
patterns that are consistent with neutrino
oscillation explanations of the atmospheric and solar data.
The absolute scale of neutrino masses is not
fixed by oscillation data
and the lightest neutrino mass may vary from 0.0-0.23 eV.}
\label{fig1}
\end{figure}

There are basically two
patterns of neutrino mass squared orderings 
consistent with the atmospheric and solar data as shown in 
Fig.\ref{fig1}.

It is clear that neutrino oscillations, which 
only depend on $\Delta m_{ij}^2\equiv m_i^2-m_j^2$, 
give no information about the absolute value of the neutrino mass squared
eigenvalues $m_i^2$ in Fig.\ref{fig1}.
Recent results from the 2df galaxy redshift survey and WMAP
indicate that $\sum m_i <0.69 {\rm eV} (95\%C.L.)$
under certain mild assumptions \cite{Elgaroy:2002bi,Pierce:2003uh}.
Combined with the solar and atmospheric oscillation data
this brackets the heaviest neutrino mass to be
in the approximate range 0.05-0.23 eV. The fact that the mass of the
heaviest neutrino is known to within an order of magnitude represents
remarkable progress in neutrino physics over recent years.

\section{CONSTRUCTING THE NEUTRINO MIXING MATRIX}

From a model building perspective the neutrino 
and charged lepton masses are given by the eigenvalues
of a complex charged lepton mass matrix
$m^E_{LR}$ and a complex symmetric neutrino Majorana matrix
$m_{LL}$, obtained by diagonalising these mass
matrices,
\beq
V^{E_L}m^E_{LR}{V^{E_R}}^{\dagger}=
\left( \begin{array}{ccc}
m_e & 0 & 0    \\
0 & m_{\mu} & 0 \\
0 & 0 & m_{\tau}
\end{array}
\right) 
\label{diag1}
\eeq
\beq
V^{\nu_L}m_{LL}{V^{\nu_L}}^T=
\left( \begin{array}{ccc}
m_1 & 0 & 0    \\
0 & m_2 & 0 \\
0 & 0 & m_3
\end{array}
\right) 
\label{diag2}
\eeq
where $V^{E_L}$, $V^{E_R}$, $V^{\nu_L}$ are unitary tranformations 
on the left-handed charged lepton fields $E_L$, right-handed charged
lepton fields $E_R$, and left-handed neutrino fields $\nu_L$
which put the mass matrices into diagonal form with real
eigenvalues. 

The neutrino mixing matrix is then constructed by
\beq
U=V^{E_L}{V^{\nu_L}}^{\dagger}
\label{MNS}
\eeq

The neutrino mixing matrix is constructed in Eq.\ref{MNS} as a product
of a unitary matrix from the charged lepton sector $V^{E_L}$
and a unitary matrix from the neutrino sector ${V^{\nu_L}}^{\dagger}$.
Each of these unitary matrices may be parametrised by its own
mixing angles and phases analagous to the parameters of $U$.
As shown in \cite{King:2002nf} the $U$ matrix can be expanded
in terms of neutrino and charged lepton mixing angles and phases
to leading order in the charged lepton mixing angles which 
are assumed to be small,
\bea
s_{23}e^{-i\delta_{23}}
& \approx &
s_{23}^{\nu_L}e^{-i\delta_{23}^{\nu_L}}
-\theta_{23}^{E_L} 
c_{23}^{\nu_L}e^{-i\delta_{23}^{E_L}}
\label{chlep23}
\\
\theta_{13}e^{-i\delta_{13}}
& \approx &
\theta_{13}^{\nu_L}e^{-i\delta_{13}^{\nu_L}}
-\theta_{13}^{E_L}c_{23}^{\nu_L}e^{-i\delta_{13}^{E_L}} \nonumber \\
& + &
\theta_{12}^{E_L}s_{23}^{\nu_L}e^{i(-\delta_{23}^{\nu_L}-\delta_{12}^{E_L})}
\label{chlep13}
\\
s_{12}e^{-i\delta_{12}} 
& \approx &
s_{12}^{\nu_L}e^{-i\delta_{12}^{\nu_L}}
+\theta_{23}^{E_L}s_{12}^{\nu_L}e^{-i\delta_{12}^{\nu_L}}
\nonumber \\
& + & \theta_{13}^{E_L}
c_{12}^{\nu_L}s_{23}^{\nu_L}e^{i(\delta_{23}^{\nu_L}-\delta_{13}^{E_L})}
\nonumber \\
& - & \theta_{12}^{E_L}
c_{23}^{\nu_L}c_{12}^{\nu_L}e^{-i\delta_{12}^{E_L}}
\label{chlep12}
\eea
Clearly $\theta_{13}$
receives important contributions not just from $\theta_{13}^{\nu_L}$,
but also from the charged lepton angles
$\theta_{12}^{E_L}$, and $\theta_{13}^{E_L}$.
In models where $\theta_{13}^{\nu_L}$ is
extremely small, $\theta_{13}$ may originate almost entirely from 
the charged lepton sector.
Charged lepton contributions could also be important in models
where $\theta_{12}^{\nu_L}=\pi /4$, since charged lepton mixing angles
may allow consistency with the LMA MSW solution.
Such effects are important for the inverted hierarchy model
\cite{King:2001ce,King:2002nf}.

\section{NEUTRINO MAJORANA MASS MATRICES}

For many (but not all) purposes it is convenient to forget about the 
division between charged lepton and neutrino mixing angles and 
work in a basis where the charged lepton mass matrix is diagonal.
Then the neutrino mixing 
angles and phases simply correspond to the neutrino ones.
In this special basis the mass matrix is given from Eq.\ref{diag2} and
Eq.\ref{MNS} as
\beq
m_{LL}=
U
\left( \begin{array}{ccc}
m_1 & 0 & 0    \\
0 & m_2 & 0 \\
0 & 0 & m_3
\end{array}
\right) 
U^T
\label{mLLnu}
\eeq
For a given assumed form of $U$ and set of neutrino masses 
$m_i$ one may use Eq.\ref{mLLnu} to ``derive'' the form of the neutrino 
mass matrix $m_{LL}$, and this results in the candidate 
mass matrices in Table \ref{table1} \cite{Barbieri:1998mq}.

\begin{table*}[htb]
{\small
\caption{Leading order low energy neutrino Majorana mass matrices
$m_{LL}$ consistent with large atmospheric and solar mixing angles,
classified according to the rate of neutrinoless double beta decay
and the pattern of neutrino masses.}
\label{table1}
\newcommand{\m}{\hphantom{$-$}}
\newcommand{\cc}[1]{\multicolumn{1}{c}{#1}}
\renewcommand{\tabcolsep}{2pc} 
\renewcommand{\arraystretch}{1.2} 
\begin{tabular}{@{}|c|c|c|}
\hline
\hline
& Type I  & Type II \\ 
 & Small $\beta \beta_{0\nu}$ & Large $\beta \beta_{0\nu}$ \\
\hline
\hline
 & & \\ 
A & $\beta \beta_{0\nu}\simlt 0.0082$ eV & \\
Normal hierarchy  & & \\
$m_1^2,m_2^2\ll m_3^2$ & 
$\left(
\begin{array}{ccc}
0 & 0 & 0 \\
0 & 1 & 1\\
0 & 1 & 1 \\
\end{array}
\right)\frac{m}{2}$ & -- \\
& & \\
\hline
 & & \\ 
B & $\beta \beta_{0\nu}\simlt
0.0082$ eV & $\beta \beta_{0\nu}\simgt 0.0085$ eV\\
Inverted hierarchy & & \\
$m_1^2\approx m_2^2\gg m_3^2$ & 
$\left(
\begin{array}{ccc}
0 & 1 & 1 \\
1 & 0 & 0\\
1 & 0 & 0 \\
\end{array}
\right)\frac{m}{\sqrt{2}}$ & 
$\left(
\begin{array}{ccc}
1 & 0 & 0 \\
0 & \frac{1}{2} & \frac{1}{2}\\
0 & \frac{1}{2} & \frac{1}{2} \\
\end{array}
\right)m$
\\
& & \\
\hline                        
 & & \\ 
C &  &
$\beta \beta_{0\nu}\simgt 0.035$ eV \\
Approximate degeneracy & & diag(1,1,1)m\\
$m_1^2\approx m_2^2\approx m_3^2$ & 
$\left(
\begin{array}{ccc}
0 & \frac{1}{\sqrt{2}} & \frac{1}{\sqrt{2}} \\
\frac{1}{\sqrt{2}} & \frac{1}{2} & \frac{1}{2}\\
\frac{1}{\sqrt{2}} & \frac{1}{2} & \frac{1}{2} \\
\end{array}
\right)m$ & 
$\left(
\begin{array}{ccc}
1 & 0 & 0 \\
0 & 0 & 1\\
0 & 1 & 0 \\
\end{array}
\right)m$\\
& & \\
\hline
\hline                        
\end{tabular}\\[2pt]
}
\end{table*}

In Table \ref{table1} the mass matrices are classified into two types:

Type I - small neutrinoless double beta decay

Type II - large neutrinoless double beta decay

They are also classified into the limiting cases consistent with the 
mass squared orderings in Fig.\ref{fig1}:

A - Normal hierarchy $m_1^2,m_2^2\ll m_3^2$

B - Inverted hierarchy $m_1^2 \approx m_2^2 \gg m_3^2$

C - Approximate degeneracy $m_1^2\approx m_2^2\approx m_3^2$

Thus according to our classification there is only one neutrino
mass matrix consistent with the normal neutrino mass hierarchy which 
we call Type IA, corresponding to the leading order neutrino masses
of the form $m_i=(0,0,m)$. For the inverted hierarchy there are two
cases, Type IB corresponding to $m_i=(m,-m,0)$ or Type IIB
corresponding to $m_i=(m,m,0)$. For the approximate degeneracy cases
there are three cases, Type IC correponding to $m_i=(m,-m,m)$
and two examples of Type IIC corresponding to either 
$m_i=(m,m,m)$ or $m_i=(m,m,-m)$. 

At present experiment allows any of the matrices in Table
\ref{table1}. In future it will be possible to uniquely specify the
neutrino matrix in the following way:

1. Neutrinoless double beta effectively measures the 11 element 
of the mass matrix $m_{LL}$ corresponding to  
\beq
\beta \beta_{0\nu}\equiv \sum_iU_{ei}^2m_i
\eeq
and is clearly capable of resolving Type I from Type II cases
according to the bounds given in Table \ref{table1} \cite{Pascoli:2002xq}.
There has been a recent claim of a signal in neutrinoless double 
beta decay correponding to $\beta \beta_{0\nu}=0.11-0.56$ eV at 95\% C.L.
\cite{Klapdor-Kleingrothaus:2001ke}.
However this claim has been criticised by two groups
\cite{Feruglio:2002af}, \cite{Aalseth:2002dt} and in turn this
criticism has been refuted \cite{Klapdor-Kleingrothaus:2002kf}. 
Since the Heidelberg-Moscow experiment has almost reached its full
sensitivity, we may have to wait for a next generation experiment
such as GENIUS \cite{petcov} which is capable 
of pushing down the sensitivity to 0.01 eV to resolve this question.

2. A neutrino factory will measure the sign of $\Delta m_{32}^2$
and resolve A from B \cite{peach}.

3. Tritium beta decay experiments are sensitive to C
since they
measure the ``electron neutrino mass'' defined by
\beq
|m_{\nu_e}|\equiv \sum_i|U_{ei}|^2|m_i|.
\eeq
For example the KATRIN \cite{weinheimer} 
experiment has a proposed sensitivity of 
0.35 eV. As already mentioned the galaxy power spectrum combined with 
solar and atmospheric oscillation data already limits the degenerate
neutrino mass to be less than about 0.6 eV, and this limit is also
expected to improve in the future. Also it is worth mentioning that
in future it may be possible to measure neutrino masses from 
gamma ray bursts using time of flight techniques in principle down to
0.001 eV \cite{Choubey:2002bh}.

Type IIB and C involve small fractional 
mass splittings $|\Delta m_{ij}^2| \ll m^2$ which are unstable 
under radiative corrections, and even the most
natural Type IC case is difficult to implement.
Types IA and IB seem to be the most natural
and later we shall focus on the normal hierarchy Type IA, 
\beq
m_{LL} \sim \left(
\begin{array}{ccc}
0 & 0 & 0 \\
0 & 1 & 1\\
0 & 1 & 1 \\
\end{array}
\right)\frac{m}{2}
\label{hier}
\eeq
However even Type IA models appear to have some remaining naturalness
problem since  
$m_3 \sim \sqrt{|\Delta m_{32}^2|}\sim 5.10^{-2}$ eV and 
$m_2 \sim \sqrt{|\Delta m_{21}^2|}\sim 7.10^{-3}$ eV, 
compared to the natural expectation $m_2 \sim m_3$.
The question may be phrased
in technical terms as one of understanding why 
the sub-determinant of the mass matrix in Eq.\ref{hier} is small:
\beq
det \left(
\begin{array}{cc}
m_{22} & m_{23}\\
m_{23} & m_{33} \\
\end{array}
\right)\ll m^2.
\label{det}
\eeq

\section{THE SEE-SAW MASS MECHANISM}

Before discussing the see-saw mechanism it is worth first reviewing
the different types of neutrino mass that are possible. So far we
have been assuming that neutrino masses are Majorana masses of the form
\beq
m_{LL}\overline{\nu_L}\nu_L^c
\label{mLL}
\eeq
where $\nu_L$ is a left-handed neutrino field and $\nu_L^c$ is
the CP conjugate of a left-handed neutrino field, in other words
a right-handed antineutrino field. Such Majorana masses are possible
to since both the neutrino and the antineutrino
are electrically neutral and so 
Majorana masses are not forbidden by electric charge conservation.
For this reason a Majorana mass for the electron would
be strictly forbidden. Majorana neutrino masses ``only''
violate lepton number conservation. 
If we introduce right-handed neutrino fields then there are two sorts
of additional neutrino mass terms that are possible. There are
additional Majorana masses of the form
\beq
M_{RR}\overline{\nu_R}\nu_R^c
\label{MRR}
\eeq
where $\nu_R$ is a right-handed neutrino field and $\nu_R^c$ is
the CP conjugate of a right-handed neutrino field, in other words
a left-handed antineutrino field. In addition there are
Dirac masses of the form
\beq
m_{LR}\overline{\nu_L}\nu_R.
\label{mLR}
\eeq
Such Dirac mass terms conserve lepton number, and are not forbidden 
by electric charge conservation even for the charged leptons and
quarks. 

In the Standard Model Dirac mass terms for charged leptons and quarks
are generated from Yukawa couplings
to a Higgs doublet whose vacuum expectation value gives the Dirac
mass term. Neutrino masses are zero in the Standard Model because
right-handed neutrinos are not present, and also 
because the Majorana mass terms in Eq.\ref{mLL}
require Higgs triplets in order to be generated at the renormalisable
level (although non-renormalisable operators can be written down.
Higgs triplets are phenomenologically
disfavoured so the simplest way to generate neutrino masses
from a renormalisable theory is to introduce right-handed neutrinos.
Once this is done then the types of neutrino mass discussed
in Eqs.\ref{MRR},\ref{mLR} (but not Eq.\ref{mLL} since we 
have not introduced Higgs triplets)
are permitted, and we have the mass matrix
\begin{equation}
\left(\begin{array}{cc} \overline{\nu_L} & \overline{\nu^c_R}
\end{array} \\ \right)
\left(\begin{array}{cc}
0 & m_{LR}\\
m_{LR}^T & M_{RR} \\
\end{array}\right)
\left(\begin{array}{c} \nu_L^c \\ \nu_R \end{array} \\ \right)
\label{matrix}
\end{equation}
Since the right-handed neutrinos are electroweak singlets
the Majorana masses of the right-handed neutrinos $M_{RR}$
may be orders of magnitude larger than the electroweak
scale. In the approximation that $M_{RR}\gg m_{LR}$ 
the matrix in Eq.\ref{matrix} may be diagonalised to 
yield effective Majorana masses of the type in Eq.\ref{mLL},
\beq
m_{LL}=m_{LR}M_{RR}^{-1}m_{LR}^T
\label{seesaw}
\eeq
This is the see-saw mechanism \footnote{For original references on the
see-saw mechanism see \cite{glashow}}. It not only generates
Majorana mass terms of the type $m_{LL}$, but also naturally makes them 
smaller than the Dirac mass terms by a factor of $m_{LR}/M_{RR}\ll 1$.
One can think of the heavy right-handed neutrinos as being integrated
out to give non-renormalisable Majorana operators suppressed
by the heavy mass scale $M_{RR}$.

In a realistic model with three left-handed neutrinos and
three right-handed neutrinos the Dirac masses $m_{LR}$
are a $3\times 3$ (complex) matrix and the heavy Majorana masses $M_{RR}$
form a separate $3\times 3$ (complex symmetric) matrix. 
The light effective Majorana
masses $m_{LL}$ are also a $3\times 3$ (complex symmetric) matrix and 
continue to be given from Eq.\ref{seesaw} which
is now interpreted as a matrix product. From a model building
perspective the fundamental parameters which must be input
into the see-saw mechanism are the Dirac mass matrix $m_{LR}$ and 
the heavy right-handed neutrino Majorana mass matrix $M_{RR}$.
The light effective left-handed Majorana mass
matrix $m_{LL}$ arises as an output according to the
see-saw formula in Eq.\ref{seesaw}. The goal of see-saw model
building is therefore to choose input see-saw matrices
$m_{LR}$ and $M_{RR}$ that will give rise to one of the successful
matrices $m_{LL}$ in Table \ref{table1}.

\section{SEQUENTIAL RIGHT HANDED NEUTRINO DOMINANCE}

With three left-handed neutrinos and
three right-handed neutrinos the Dirac masses $m_{LR}$
are a $3\times 3$ (complex) matrix and the heavy Majorana masses $M_{RR}$
form a separate $3\times 3$ (complex symmetric) matrix. 
The light effective Majorana
masses $m_{LL}$ are also a $3\times 3$ (complex symmetric) matrix and 
continue to be given from Eq.\ref{seesaw} which
is now interpreted as a matrix product. From a model building
perspective the fundamental parameters which must be input
into the see-saw mechanism are the Dirac mass matrix $m_{LR}$ and 
the heavy right-handed neutrino Majorana mass matrix $M_{RR}$.
The light effective left-handed Majorana mass
matrix $m_{LL}$ arises as an output according to the
see-saw formula in Eq.\ref{seesaw}. The goal of see-saw model
building is therefore to choose input see-saw matrices
$m_{LR}$ and $M_{RR}$ that will give rise to one of the successful
matrices $m_{LL}$ in Table \ref{table1}.

We now show how the input see-saw matrices can be simply chosen to 
give the Type IA matrix in Eq.\ref{hier}, with the property of a
naturally small sub-determinant in Eq.\ref{det} using a mechanism
first suggested in \cite{King:1998jw}.
The idea was developed in \cite{King:1999cm} where it was called
single right-handed neutrino dominance (SRHND) . SRHND was first successfully
applied to the LMA MSW solution in \cite{King:2000mb}.  

The SRHND mechanism is most simply described 
assuming three right-handed neutrinos
in the basis where the right-handed neutrino mass matrix is diagonal
although it can also be developed in other bases 
\cite{King:1999cm,King:2000mb}. In this basis we write the input
see-saw matrices as
\begin{equation}
M_{RR}=
\left( \begin{array}{ccc}
X' & 0 & 0    \\
0 & X & 0 \\
0 & 0 & Y
\end{array}
\right) 
\label{seq1}
\end{equation}
\begin{equation}
m_{LR}=
\left( \begin{array}{ccc}
a' & a & d    \\
b' & b & e \\
c' & c & f
\end{array}
\right) 
\label{dirac}
\end{equation}
In \cite{King:1998jw} it was suggested that one of the right-handed neutrinos 
may dominante the contribution to $m_{LL}$ if it is lighter than
the other right-handed neutrinos. 
The dominance condition was subsequently generalised to 
include other cases where the right-handed neutrino may be
heavier than the other right-handed neutrinos but dominates due to its larger
Dirac mass couplings \cite{King:1999cm}.
In any case the dominant neutrino may be taken to be the third one 
without loss of generality. Assuming SRHND then Eqs.\ref{seesaw},
\ref{seq1}, \ref{dirac} give, retaining only the leading dominant
right-handed neutrino contributions proportional to $1/Y$,
\beq
m_{LL}
\approx
\left( \begin{array}{ccc}
\frac{d^2}{Y}
& \frac{de}{Y}
& \frac{df}{Y}    \\
.
& \frac{e^2}{Y} 
& \frac{ef}{Y}    \\
. & . & \frac{f^2}{Y} 
\end{array}
\right)
\label{one}
\eeq
If the Dirac mass couplings satisfy the condition 
$d\ll e\approx f$ \cite{King:1998jw}
then the matrix in Eq.\ref{one} resembles the Type IA matrix in
Eq.\ref{hier}, and furthermore has a naturally small sub-determinant as in
Eq.\ref{det}. The neutrino mass
spectrum consists of one neutrino with mass $m_3\approx (e^2+f^2)/Y$
and two approximately
massless neutrinos \cite{King:1998jw}. The atmospheric angle is
$\tan \theta_{23} \approx e/f$ \cite{King:1998jw}.
It was pointed out that small perturbations from
the sub-dominant right-handed neutrinos can then lead to a small
solar neutrino mass splitting \cite{King:1998jw}.

It was subsequently shown how to
account for the LMA MSW solution with a large solar angle
\cite{King:2000mb} by careful consideration of the sub-dominant contributions. 
One of the examples considered in \cite{King:2000mb} is when the 
right-handed neutrinos dominate sequentially,
\beq
\frac{|e^2|,|f^2|,|ef|}{Y}\gg
\frac{|xy|}{X} \gg
\frac{|x'y'|}{X'}
\label{srhnd}
\eeq
where $x,y\in a,b,c$ and $x',y'\in a',b',c'$.
Assuming SRHND with sequential sub-dominance as in
Eq.\ref{srhnd}, then Eqs.\ref{seesaw}, \ref{seq1}, \ref{dirac} give
\beq
m_{LL}
\approx
\left( \begin{array}{ccc}
\frac{a^2}{X}+\frac{d^2}{Y}
& \frac{ab}{X}+ \frac{de}{Y}
& \frac{ac}{X}+\frac{df}{Y}    \\
.
& \frac{b^2}{X}+\frac{e^2}{Y} 
& \frac{bc}{X}+\frac{ef}{Y}    \\
.
& .
& \frac{c^2}{X}+\frac{f^2}{Y} 
\end{array}
\right)
\label{mLL2}
\eeq
where the contribution from the first right-handed neutrino may be 
neglected according to Eq.\ref{srhnd}. 
This was shown to lead to a full neutrino mass hierarchy
\beq
m_1^2\ll m_2^2\ll m_3^2
\eeq
and, ignoring phases, the solar angle only depends
on the sub-dominant couplings and is given by 
$\tan \theta_{12} \approx a/(c_{23}b-s_{23}c)$ \cite{King:2000mb}.
The simple requirement for large solar angle is then $a\sim b-c$ 
\cite{King:2000mb}.

Including phases the neutrino masses
are given to leading order in $m_2/m_3$ by diagonalising
the mass matrix in Eq.\ref{mLL2} using the analytic proceedure
described in \cite{King:2002nf},
\bea
m_1 & \sim & O(\frac{x'y'}{X'}v_2^2) \label{m1} \\
m_2 & \approx &  \frac{|a|^2}{Xs_{12}^2} v_2^2 \label{m2} \\
m_3 & \approx & \frac{|e|^2+|f|^2}{Y}v_2^2 
\label{m3}
\eea
where $v_2$ is a Higgs vacuum expectation value (vev) associated with
the (second) Higgs doublet that couples to the neutrinos
and $s_{12}=\sin \theta_{12}$ given below. 
Note that with SD each neutrino mass is generated
by a separate right-handed neutrino, and the sequential dominance condition
naturally results in a neutrino mass hierarchy $m_1\ll m_2\ll m_3$.
The neutrino mixing angles are given to leading order in $m_2/m_3$ by, 
\bea
\tan \theta_{23} & \approx & \frac{|e|}{|f|} \label{23}\\
\tan \theta_{12} & \approx &
\frac{|a|}
{c_{23}|b|
\cos(\tilde{\phi}_b)-
s_{23}|c|
\cos(\tilde{\phi}_c)} \label{12} \\
\theta_{13} & \approx &
e^{i(\tilde{\phi}+\phi_a-\phi_e)}
\frac{|a|(e^*b+f^*c)}{[|e|^2+|f|^2]^{3/2}}
\frac{Y}{X}
\label{13}
\eea
where we have written some (but not all) complex Yukawa couplings as
$x=|x|e^{i\phi_x}$. The phase $\delta$
is fixed to give a real angle
$\theta_{12}$ by,
\beq
c_{23}|b|
\sin(\tilde{\phi}_b)
\approx
s_{23}|c|
\sin(\tilde{\phi}_c)
\label{chi1}
\eeq
where 
\bea
\tilde{\phi}_b &\equiv & 
\phi_b-\phi_a-\tilde{\phi}+\delta, \nonumber \\ 
\tilde{\phi}_c &\equiv & 
\phi_c-\phi_a+\phi_e-\phi_f-\tilde{\phi}+\delta
\label{bpcp}
\eea
The phase $\tilde{\phi}$
is fixed to give a real angle
$\theta_{13}$ by,
\beq
\tilde{\phi} \approx  \phi_e-\phi_a -\phi_{\rm COSMO}
\label{phi2dsmall}
\eeq
where
\beq
\phi_{\rm COSMO}=\arg(e^*b+f^*c). 
\label{lepto0}
\eeq
is the leptogenesis phase 
corresponding to the interference diagram involving the
lightest and next-to-lightest right-handed neutrinos \cite{King:2002nf}.

\section{LIGHT OR HEAVY SEQUENTIAL DOMINANCE?}
Assuming sequential dominance described in the previous section,
there is still an ambiguity regarding the mass ordering of the 
heavy Majorana right-handed neutrinos. There are two extreme possibilities
called heavy sequential dominance (HSD) and light sequential dominance
(LSD). In HSD the dominant right-handed
neutrino (always denoted by Majorana mass $Y$) is the heaviest,
\beq
X'\ll X \ll Y
\eeq
Then assuming that the 33 element of the neutrino Yukawa matrix is
of order unity, this leads to a ``lop-sided'' Yukawa matrix,
since $e\sim f \sim 1$,
\begin{equation}
M^{HSD}_{RR}=
\left( \begin{array}{ccc}
X' & 0 & 0    \\
0 & X & 0 \\
0 & 0 & Y
\end{array}
\right) 
\label{hsdmaj}
\end{equation}
\begin{equation}
{Y_{\nu}}^{HSD}_{LR}=
\left( \begin{array}{ccc}
a' & a & d    \\
b' & b & e \\
c' & c & f
\end{array}
\right) 
\label{hsdyuk}
\end{equation}

On the other hand in LSD, the dominant right-handed neutrino of mass
$Y$ is by definition the lightest one,
\beq
Y\ll X \ll X'
\eeq
Then still assuming that the 33 element of the neutrino Yukawa matrix is
of order unity, this leads to a ``quark-like'' Yukawa matrix,
since in this case $e\sim f \ll 1$
consistent with a symmetrical structure with no large off-diagonal elements,
after reordering the right-handed neutrinos,
\begin{equation}
M^{LSD}_{RR}=
\left( \begin{array}{ccc}
Y & 0 & 0    \\
0 & X & 0 \\
0 & 0 & X'
\end{array}
\right) 
\label{lsdmaj}
\end{equation}
\begin{equation}
{Y_{\nu}}^{HSD}_{LR}=
\left( \begin{array}{ccc}
d & a & a'    \\
e & b & b' \\
f & c & c'
\end{array}
\right) 
\label{lsdyuk}
\end{equation}

Note that in LSD, the right-handed neutrino of mass $X'$ is irrelevant
for neutrino masses and mixings, as well as leptogenesis.
For all practical purposes, the LSD model reduces to an effective
two right-handed neutrino model. 

\section{LEPTON FLAVOUR VIOLATION IN THE CMSSM WITH SEQUENTIAL DOMINANCE}
At leading order in a mass insertion approximation 
\footnote{For a complete list of references on lepton flavour
violatio see \cite{Blazek:2002wq}.} 
the branching fractions of LFV processes are given by
\beq
{\rm BR}(l_i \rightarrow l_j \gamma)\approx 
        \frac{\alpha^3}{G_F^2}
        f(M_2,\mu,m_{\tilde{\nu}}) 
        |m_{\tilde{L}_{ij}}^2|^2  \tan ^2 \beta
    \label{eq:BR(li_to_lj)}
\eeq
where $l_1=e, l_2=\mu , l_3=\tau$,
and where the off-diagonal slepton doublet mass squared is given 
in the leading log approximation (LLA) by
\beq
m_{\tilde{L}_{ij}}^{2(LLA)}
\approx -\frac{(3m_0^2+A_0^2)}{8\pi ^2}C_{ij}
\label{lla}
\eeq
where the leading log coefficients relevant for 
$\mu \rightarrow e\gamma$ and $\tau \rightarrow \mu \gamma$
are given approximately as
\bea
C_{21} & = & ab\ln \frac{M_U}{X} +de\ln \frac{M_U}{Y} \nonumber \\
C_{32} & = & bc\ln \frac{M_U}{X} +ef\ln \frac{M_U}{Y}
\label{C2131}
\eea

We have performed a global analysis of LFV in the constrained
minimal supersymmetric standard model (CMSSM) for the case
of sequential dominance, focussing on the two cases
of HSD and LSD \cite{Blazek:2002wq}.
We parametrise the matrices \cite{Blazek:2002wq} in a quite general
way consistent with sequential dominance. The numerical results we
show here are for a particular case of HSD and LSD defined below
\bea
M^{HSD}_{RR}&=&
\left( \begin{array}{ccc}
- & 0 & 0    \\
0 & \lambda^{3} & 0 \\
0 & 0 & 1
\end{array}
\right) 3.10^{14}GeV \label{HSD1} \\
Y^{\nu HSD}_{LR}&=&
\left( \begin{array}{ccc}
- & a_{12}\lambda^2 & 0    \\
- & a_{22}\lambda^2 & 1 \\
- & a_{32}\lambda^2 & 1
\end{array}
\right)
\label{HSD2}
\eea
\bea
M^{LSD}_{RR}&=&
\left( \begin{array}{ccc}
\lambda^{6} & 0 & 0    \\
0 & \lambda^{3} & 0 \\
0 & 0 & \gg 1
\end{array}
\right) 3\times 10^{14} GeV \label{LSD1} \\
Y^{\nu LSD}_{LR}&=&
\left( \begin{array}{ccc}
0 & a_{12}\lambda^{2} & -    \\
\lambda^{3} & a_{22}\lambda^{2} & - \\
\lambda^{3} & a_{32}\lambda^{2} & 1
\end{array}
\right)
\label{LSD2}
\eea
where $a_{ij}$ are order unity coefficients, 
$\lambda=\sqrt{\Delta m_{21}^2/\Delta m_{32}^2}\approx 0.15$.

\begin{figure}
\vspace*{13pt}
        \mbox{\epsfig{figure=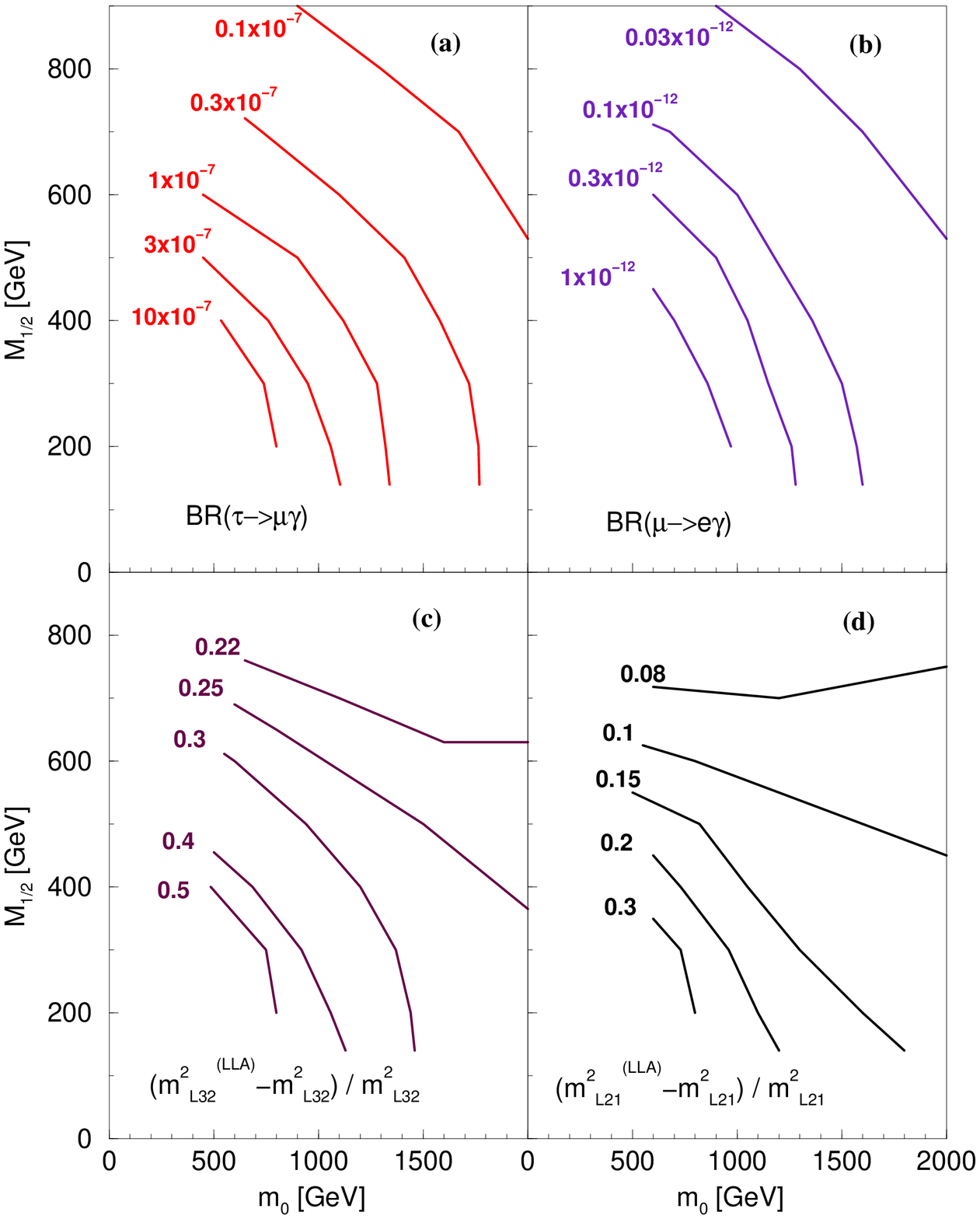,width=14.0cm}}
\vspace*{1.4truein}		
\caption{The upper panels show the predictions for the branching fraction 
for (a) $\tau \rightarrow \mu \gamma$ and (b) $\mu \rightarrow e\gamma$
for HSD using the parameters described in the text.
The lower panels (c), (d)
show the fractional error $\Delta_{ij}$, defined in the
text, that would be
made in calculating the off-diagonal slepton masses if the leading log
approximation had been used instead of the exact calculation.}
\label{hsd}
\end{figure}

\begin{figure}
\vspace*{13pt}
        \mbox{\epsfig{figure=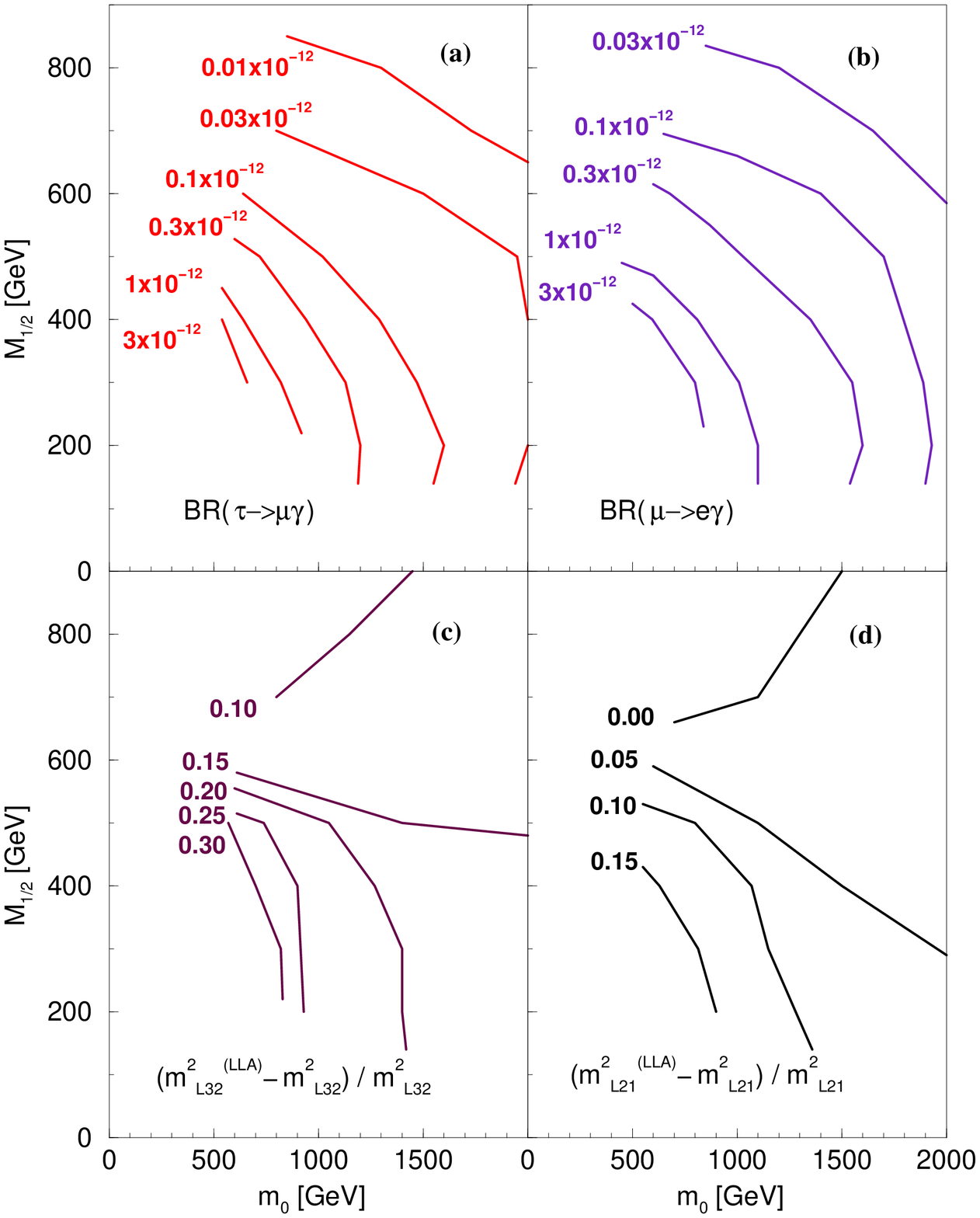,width=14.0cm}}
\vspace*{1.4truein}		
\caption{The upper panels show the predictions for the branching fraction 
for (a) $\tau \rightarrow \mu \gamma$ and (b) $\mu \rightarrow e\gamma$
for LSD using the parameters described in the text.
The lower panels (c), (d)
show the fractional error $\Delta_{ij}$, defined in the
text, that would be
made in calculating the off-diagonal slepton masses if the leading log
approximation had been used instead of the exact calculation.}
\label{lsd}
\end{figure}

\begin{figure}
\vspace*{13pt}
        \mbox{\epsfig{figure=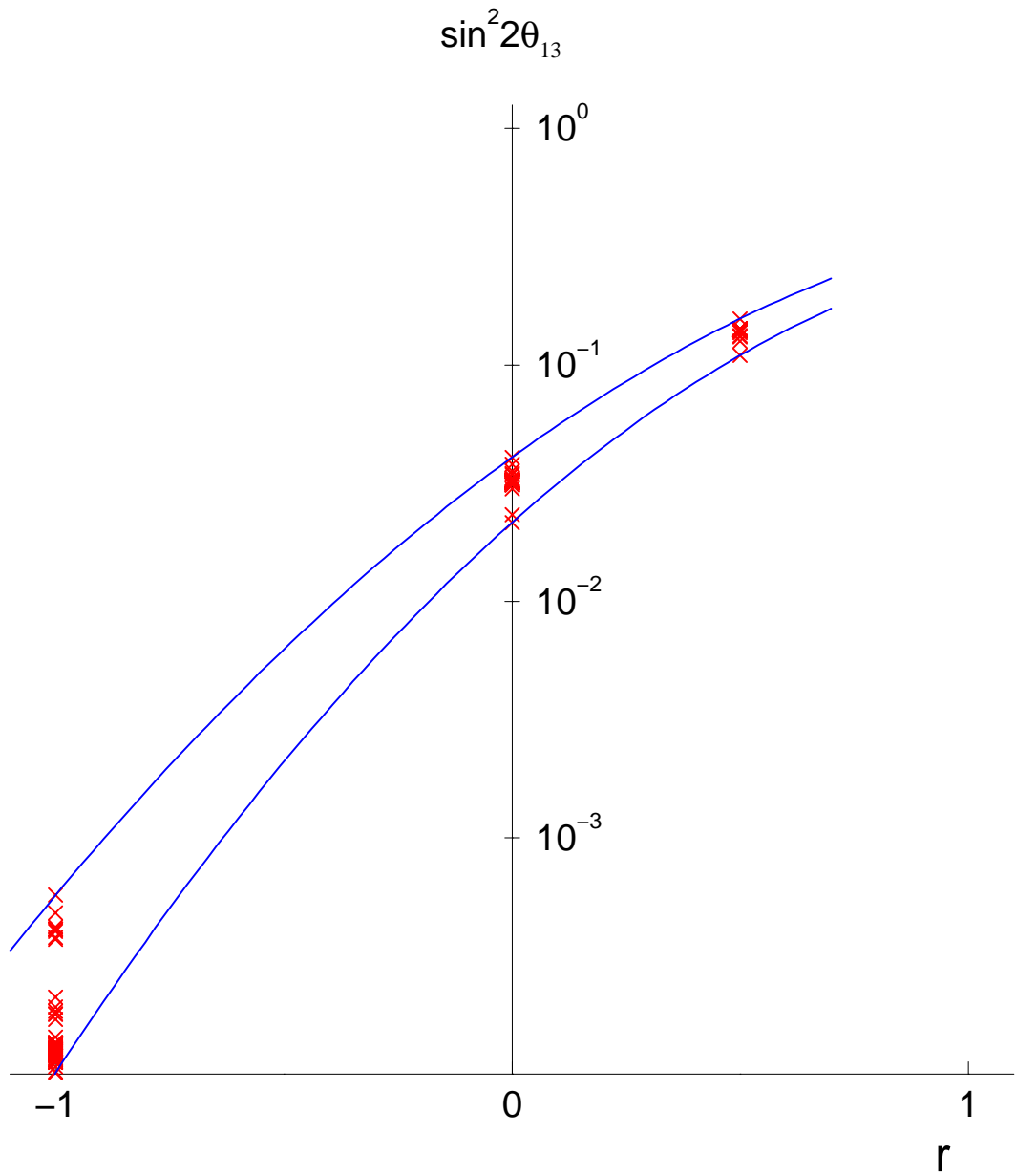,width=14.0cm}}
\vspace*{1.4truein}		
\caption{The prediction of $\sin^22\theta_{13}$ as a function of 
the ratio of subdominant Yukawa couplings $r=a_{32}/a_{22}$
for the HSD class of models.} 
\label{theta}
\end{figure}

In Figure \ref{hsd} we show results for HSD for $\tan \beta=50$,
and $r=a_{32}/a_{22}=-1$. 
The results show a large rate for
$\tau \rightarrow \mu \gamma$ which is the characteristic expectation
of lop-sided models in general \cite{Blazek:2001zm} and HSD in particular.
We also show the error incurred
if the LLA were used (our results are based on an exact calculation).
In Figure \ref{lsd} we show results for LSD for $\tan \beta=50$,
and $r=a_{32}/a_{22}=-1$.
The results show a much smaller rate for
$\tau \rightarrow \mu \gamma$, and also the error incurred
if the LLA were used (our results are based on an exact calculation).
In Figure \ref{theta} we show that the 13 mixing angle is controlled
by a ratio of subdominant Yukawa couplings, 
$r=a_{32}/a_{22}$ and therefore cannot be
predicted in general, although in particular models this ratio
may be fixed by the theory.

\section{LEPTOGENESIS AND ITS POSSIBLE RELATION TO CP VIOLATION MEASURED IN FUTURE NEUTRINO OSCILLATION EXPERIMENTS}

Is there a link between the CP violation required for leptogenesis,
and the phase $\delta$ measurable in accurate neutrino oscillation
experiments? In general the answer would seem to be no, however
in certain classes of models the answer can be yes. 
For example in sequential dominance we find that for HSD there is 
no link since the lightest right-handed neutrino of mass $X'$
is quite relevant for leptogenesis, but completely
irrelevant for the determining the neutrino angles and phases.
However, in LSD there may be a link since in this case the lightest
right-handed neutrino is also the dominant one, and so plays
an important part in both leptogenesis and in determining the
neutrino mixings and phases. Moreover in LSD the heaviest
right-handed neutrino of mass $X'$ is irrelevant for both
leptogenesis and neutrino mixings and phases, so the model
effectively reduces to a two right-handed neutrino model
\cite{King:2002nf}.

The details of this have been recently worked out for the LSD class
of models in which the neutrino matrices are as in 
Eqs.\ref{lsdmaj},\ref{lsdyuk} assuming the sequential dominance
condition in Eq.\ref{srhnd}, and in addition assuming that $d=0$
which corresponds in this case to a 11 texture zero
\cite{King:2002qh}.
Although the class of model looks quite specialised,
it is in fact extremely well motivated since it 
allows the neutrino Yukawa matrix to have the same universal form
as the quark Yukawa matrices 
consistent with $SO(10)$ for example, where the
11 texture zero arises naturally  \cite{King:2001uz}.

Returning to Eq.\ref{lepto0}, it may be expressed as
\beq
\tan \phi_{\rm COSMO} \approx 
\frac{|b|s_{23}s_2+|c|c_{23}s_3}{|b|s_{23}c_2+|c|c_{23}c_3}.
\label{phi121}
\eeq
Inserting $\tilde{\phi}$ in Eq.\ref{phi2dsmall}
into Eqs.\ref{chi1},\ref{bpcp},
\bea
&&c_{23}|b|\sin(\eta_2+\phi_{\rm COSMO}+\delta) \nonumber \\
& \approx &
s_{23}|c|
\sin(\eta_3+\phi_{\rm COSMO} +\delta).
\label{chi12}
\eea
Eq.\ref{chi12} may be expressed as
\beq
\tan (\phi_{\rm COSMO}+\delta) \approx 
\frac{|b|c_{23}s_2-|c|s_{23}s_3}{-|b|c_{23}c_2+|c|s_{23}c_3}
\label{phi12del}
\eeq
where we have written $s_i=\sin \eta_i, c_i=\cos \eta_i$
where
\beq
\eta_2\equiv \phi_b-\phi_e, \ \ \eta_3\equiv \phi_c-\phi_f
\label{eta}
\eeq
are invariant under a charged lepton phase transformation.
The reason that the see-saw parameters only
involve two invariant phases $\eta_2, \eta_3$ rather than the usual six
is due to the LSD 
assumption which has the effect of decoupling the  
heaviest right-handed neutrino, which removes three phases, 
together with the assumption
of a 11 texture zero, which removes another phase.

Eq.\ref{phi12del} shows that
$\delta$ is a function of 
the two see-saw phases
$\eta_2 , \eta_3$ that also determine $\phi_{\rm COSMO}$ in Eq.\ref{phi121}.
If both the phases $\eta_2 , \eta_3$ are zero,
then both $\phi_{\rm COSMO}$ and $\delta$ are necessarily zero.
This feature is absolutely crucial. It means that,
barring cancellations, measurement of a non-zero value for 
the phase $\delta$ at a neutrino factory will be a signal of a
non-zero value of the leptogenesis phase $\phi_{\rm COSMO}$.
We also find the remarkable result
\beq
|\phi_{\rm COSMO}|=|\phi_{\beta \beta 0\nu}|.
\label{remarkable2}
\eeq
where $\phi_{\beta \beta 0\nu}$ is the phase which enters the rate
for neutrinoless double beta decay \cite{King:2002qh}.

To conclude, we have discussed the relation between leptogenesis 
and the MNS phases in the LSD class of models defined by
Eqs.\ref{lsdmaj},\ref{lsdyuk},\ref{srhnd} with the additional
assumption of a 11 texture zero.
Although the class of model looks quite specialised,
it is in fact extremely well motivated since it 
allows the neutrino Yukawa matrix to have the same universal form
as the quark Yukawa matrices 
consistent with $SO(10)$ for example, where the
11 texture zero arises naturally  \cite{King:2001uz}.
The large neutrino mixing angles and neutrino mass hierarchy
then originate naturally
from the sequential dominance mechanism
without any fine tuning \cite{King:2000mb}.
Within this class of models we have shown that
the two see-saw phases $\eta_2$, $\eta_3$ 
are related to $\delta $ and $\phi_{\rm COSMO}$
according to Eqs.\ref{phi121},\ref{phi12del}.
Remarkably, the leptogenesis phase is predicted to be
equal to the neutrinoless double beta decay phase as in
Eq.\ref{remarkable2}. Since the heaviest right-handed
neutrino of mass $X'$ is irrelevant for both leptogenesis and
for determining the neutrino masses and mixings, the model
reduces effectively to one involving only two right-handed neutrinos
\cite{King:2000mb}. 

In this talk I have demonstrated that the
physics of neutrino mass, flavour and CP violation are all closely
linked. When the information from the neutrino sector is combined with
that from the quark sector, including ideas of family symmetry and 
unification \cite{King:2001uz}, it is just possible that it may be
enough to unlock the whole mystery of flavour.

\section{Acknowledgements}
I would like to thank Milla Baldo Ceolin for her kind hospitality
at the X International Workshop on "Neutrino Telescopes".

\end{document}